# Time-to-event estimands and loss to follow-up in oncology in light of the estimands framework

*Jonathan Siegel (Bayer)), Hans-Jochen Weber (Novartis), Stefan Englert (AbbVie), Feng Liu (Marengo Therapeutics), & on behalf of the Industry Working Group on Estimands in Oncology*

Short Title: Time-to-event estimands in oncology

**Abstract**

Time-to-event estimands are central to many oncology clinical trials. The estimand framework (addendum to the ICH E9 guideline) calls for precisely defining the treatment effect of interest to align with the clinical question of interest and requires predefining the handling of intercurrent events that occur after treatment initiation and either preclude the observation of an event of interest or impact the interpretation of the treatment effect. We discuss a practical problem in clinical trial design and execution, i.e. in some clinical contexts it is not feasible to systematically follow patients to an event of interest. Loss to follow-up in the presence of intercurrent events can affect the meaning and interpretation of the study results. We provide recommendations for trial design, stressing the need for close alignment of the clinical question of interest and study design, impact on data collection and other practical implications. When patients cannot be systematically followed, compromise may be necessary to select the best available estimand that can be feasibly estimated under the circumstances. We discuss the use of sensitivity and supplementary analyses to examine assumptions of interest.

# Table of Contents

## 1. Introduction

With the release of the addendum to the ICH E9 guidance[1], the estimand framework is being applied widely in the design and analysis of clinical trials. The estimand framework provides terminology to clearly describe what a clinical trial is intending to assess and consequently, which data need to be collected and how these data are to be analysed. This paper is a product of the Pharmaceutical Industry Working Group on Estimands in Oncology's Censoring Mechanisms Subteam.

For an overview of the causal inference framework, see Pearl[2]. For an overview of the estimand framework in clinical trials generally, see e.g. Akacha et al.[3]; Ratich et al.[4]; and Malinckrodt et al.[5] and for estimands in time-to-event endpoints, see Rufibach[6].

In oncology trials, assessing the treatment effect of time to event endpoints is typically of high interest and referenced in health authority guidelines. In the past, intercurrent events were frequently addressed by censoring rules in traditional censoring tables. As we will discuss, the principles underlying the estimands framework, however, cast significant doubt on past use of censoring as a means of handling intercurrent events.

In this paper we consider that while it is often desirable to follow all patients to the clinical event of interest, in some cases it is not possible or not feasible to follow patients through and beyond particular intercurrent events. Lack of follow-up in such cases requires careful definition and interpretation of the relevant estimand. We identify a number of real-world clinical situations where consistent follow-up may not be feasible and discuss strategies to address them. In addition to attempting to address the inability to follow patients until the clinical event of interest, we provide general strategies and recommendations for clarifying the clinical question of interest, defining estimands, selecting strategies for addressing intercurrent events, collecting data, censoring, and defining sensitivity analyses in the context of time-to-event endpoints. We also discuss more general cases where the originally desired estimand might be infeasible in a particular context, and discuss ways to develop alternatives.

## 2.    Background

A recurring problem in oncology clinical trials with survival estimands is situations where all or a substantial fraction of patients cannot feasibly be followed beyond an intercurrent event when follow-up is mandated by the chosen strategy. Examples include:

- Patients can never be followed beyond terminal events, such as death. As mortality is, unfortunately, pervasive in cancer, appropriately accounting for death requires careful consideration in virtually all estimands in an oncology trial.
- Within some open-label trials or trials with functional unblinding due to side effects, large fractions of patients randomized to placebo have left the trial shortly after randomization. An example is the Checkmate-037 trial, where 37% of patients randomized to placebo immediately went off treatment and took other immunotherapies as subsequent therapy[7]. See generally Manitz (2021)[8].
- Although it is increasingly possible to continue clinic visits and assessments beyond treatment discontinuation, it is not always possible to do so beyond events such as radiological progression. Patients might for example move to another trial. Although devices for recording Patient Reported Outcomes, patient diaries, etc. are increasing the scope of home data collection, many kinds of assessments can still only be collected reliably at clinic visits. When clinic visits cease, assessments dependent on clinic visits also cease.

As Unkel (2019) explained regarding adverse events:

> A major challenge in oncology is the treatment change after progression of the disease, where in many cases patients enter a subsequent clinical study, eg, in malignant melanoma where about 4 years ago, the only treatment option was dacarbacin, and most patients entered clinical trials after progression. These studies were under evaluation by the HTA bodies recently due to the time gap between study conduct and marketing authorization. However, in most studies, a patient is not allowed to enter a new clinical study, if they are still participants of the prior study. As a consequence, they need to withdraw consent for the first study to enter the next. Therefore, AEs cannot be collected after progression for the first study in general.[9]

While practices regarding permitting entry to subsequent studies have sometimes been more flexible, what Unkel described for AEs applies more generally for efficacy indicators requiring clinic visits to measure.

In the past, time-to-event methods have simply censored patients at the end of all assessments, often without regard to whether such censoring is informative, or whether patients might have ended assessments as a result of an intercurrent event. In addition, past regulatory guidance, such as the 2007 FDA Cancer Endpoint Guidance[10], mandated censoring for certain intercurrent events such as subsequent therapy in analyses of PFS. The current 2018 FDA Cancer Endpoint Guidance[11] retained this approach, although making it optional.

### 3. The importance of clinical questions

In designing trials under the estimands framework, practitioners should begin with clarifying the clinical questions of interest. The clinical question, clearly defined, will in particular lead to determining the appropriate strategy for handling intercurrent events, and the appropriate estimation methods to implement those strategies. As we discuss in this paper, in some cases the desired clinical question cannot feasibly be answered in a particular study context, as it may simply not be possible to meet some of the assumptions or conditions required to answer the research questions of most interest. In a time-to-event trial, it may not be feasible either to follow patients through and beyond particular intercurrent events, or to assume that the effect of the intercurrent event can be incorporated into the treatment effect. When feasibility issues occur, study design teams incorporating multiple fields of expertise (clinical research, clinical practice, statistics, and study operations) will need to work as cooperative team, using an iterative process, to formulate the best feasible question that can be answered under the circumstances. An effective trial may need to be a compromise between what is desired and what is feasible. The desired clinical question of interest should, however, always comes first. It must guide practice and illuminate any compromises that need to be made.

In presenting our work, we make a simplification. Our approach is laid out assuming that researchers will typically start out interested in questions that can be addressed by assuming a lab experiment model, which generally corresponds, as we discuss below, to a treatment policy strategy.. Under this simplification, an initial lab-metaphor clinical question needs to be adjusted when either the subject-matter renders it inappropriate or trial conditions renders it infeasible. This simplification reflects the the influence of the intent-to-treat (ITT) principle in the 1998 ICH E9 guidance.[12]

In many contexts alternative approaches are currently highly controversial or not accepted at all by regulators. Accepting a treatment policy strategy as a default, broken when intercurrent events render it infeasible, lets us help frame specific arguments for alternative strategies where we think an argument is warranted. Our approach is intended not merely to facilitate trial design, but also to help guide industry-regulatory discussions.  When intercurrent events jeopardize a treatment policy strategy, other strategies will need to be considered. We recognize that treating a treatment policy approach as a default represents a simplification specific to and designed to reflect the needs of the pharmaceutical industry in addressing the perspectives of regulators. We also recognize that while we hope it proves a useful model, this simplification does not always hold. There are situations where the current presumptive strategy in pharmaceutical practice is not a treatment policy strategy. This is particularly true in safety, for example.[9] This paper's general approach may be more appropriate for efficacy.

# 4. Navigating theory and practice: Some history, context, and terminology

The authors seek to provide a theoretical approach consistent with classical survival analysis and the causal estimands framework, in which an estimand describes the realm of the population of inference and is not affected by what happens with the sample during the course of a trial, and also to discuss practical issues in how to discuss the effect of loss to follow-up on a proposed estimand or strategy at the design stage, including how to evaluate clinical trials after the fact. We are proposing terminology that may help address both needs. In some cases industry use would better conform to classical theory. In other cases, however, industry practice has special needs that may require supplementary concepts to describe.

**The meaning and role of censoring**

The 2007 FDA Cancer Endpoint guidance,[10] in effect until replaced in 2018,[11] mandated the use of censoring tables and censoring for a variety of circumstances, including censoring for subsequent therapy for PFS. The pharmaceutical clinical trials industry has accordingly become accustomed to elaborate censoring tables detailing censoring for a variety of conditions.

In both classical survival analysis literature,[13,14,15,16] and in literature on survival analysis within the causal estimands framework,[17,18,19,20] censoring has a narrower role. It is merely a technique for addressing lack of further data without an event having occurred. And as such it lies entirely within the realm of the estimator. While censoring tables have become endemic to pharmaceutical-industry time-to-event clinical trials, in both classical survival analysis and the causal estimands framework there are no censoring tables. The only basis for censoring would be lack of further data without the event of interest having occurred, after all intercurrent events with specific strategies and implementation methods have been accounted for.

The ICH E9 (R1) Guidance was written to be generally consistent with the causal estimands framework. It refers to some of the framework's concepts and terms. The Guidance does not, however, explicitly endorse specific statistical theories.[1] To date, industry adaption of the estimands framework has been partial. We expect that, regardless of their theoretical merits, censoring tables will likely continue in practice, and will continue to need to be interpreted, for the foreseeable future.

In this paper, balancing between the causal estimands viewpoint of much of the academic survival analysis community and traditional pharmaceutical industry definitions and practices, we refer to traditional censoring practices of the type used in traditional censoring tables by putting "censoring" in quotes, except where otherwise noted. And except where otherwise noted, we use censoring without quotation marks only to refer to catch-all adjustment for the end of assessments without an event of interest as part of estimation technique, without censoring "for" anything specific.

**Estimand and implied estimand**

One of the features of regulatory clinical trials, and oncology trials in particular, is that late-phase studies are often planned and launched based on the results of a Phase I or small-sample Phase IIa study. Survival and hazards distributions, censoring patterns, and key intercurrent events may not be fully planned for and may not be known. For this reason, it will continue to be necessary to interpret trials after the fact not only to understand how reliable their underlying assumptions turned out to be, but also to interpret their meaning in light of what is known post-hoc.

The estimands framework, on the other hand, stresses pre-planning and addressing issues at the design stage, which tends to require knowing key assumptions in advance, at study design. The fact that this may not always be possible in practice given the pressures on oncology trialists to deliver early represents a potential gap between theoretician and practitioner.

The framework maintains a sharp distinction between estimand, the realm of population of inference, and estimator, the realm of the study sample and trial conduct. To maintain this distinction while enabling practitioners to talk about how unexpected study conduct can change the meaning and interpretation of the results, we distinguish between an *estimand* and an *implied estimand*. The concept of implied estimand was introduced by Rufibach[6] to address the post-hoc interpretation of the data as distinct from the estimand planned at study design. Accordingly:

An **estimand** (clarified only with respect to this one issue) represents a clinical question formulated at trial design.

An **implied estimand**[6] represents the best post-hoc interpretation of the data, "reverse engineering" the analysis method, as impacted by trial conduct, data rules, etc. It can be different from the originally specified estimand.

**Process termination**

Our next set of terminology addresses the impact of deletion of events on an estimand, or implied estimand. One of the key points of our paper, discussed further below, is that under some circumstances, loss to follow-up occurring in a study can not only result in bias with respect to the original estimand, but can result in an implied estimand, a best post-hoc interpretation of the data, that is different from the one originally intended. In particular, we discuss circumstances where on-trial conduct defeats use of a treatment policy strategy, the most common strategy used in regulatory pharmaceutical time-to-event trials.

We note that isolated informative censoring, while potentially introducing bias, will not generally change the overall interpretation of the results. For that reason, we limit our terms to pervasive or systematic patterns of censoring, which as we discuss below may indeed do so. In doing this, we acknowledge introducing an element of subjective judgment. We are not proposing a test to determine exactly when the degree of loss to follow-up is sufficiently pervasive to change the interpretation. Matters of interpretation tend not to be so precise, and new methods might permit a reliable estimate with less follow-up than in the past.

We use the term *process termination* to refer to the systematic or pervasive deletion of events from an underlying process prior to the observation of the event of interest. *Natural process termination* is the systematic deletion of events from the underlying real-world process, the result of terminal events such as death. Natural process termination, being a property of the process modelled, is necessarily a property of the estimand. *Artificial process termination*, on the other hand, is the deletion of subsequent events from the measurement process without deleting them from the underlying real-world process. Artificial process termination can occur by means of traditional "censoring" tables, or by pervasive patient drop-out.

**Process termination** is the systematic or pervasive deletion of events from the measurement process prior to the observation of the event of interest, whether due to the underlying real-world process or due to an artifact of measurement.

**Natural process termination** occurs when an event in the underlying real-world process being observed, a terminal event, systematically results in deletion of events from the underlying estimand. Natural process termination is a property of the estimand.

**Artificial process termination** occurs when subsequent events are systematically or pervasively deleted from the measurement process without deletion from the underlying real-world process. Artificial process termination is a property of the estimator.

## 5. Time to event endpoints in oncology

Substantial progress has been made over recent years to understand, diagnose and to treat cancer. Life expectancy has increased in many indications,[21] however, prolonging life remains the key objective in cancer patients. Therefore, overall survival is the gold standard endpoint and well acknowledged by Health Authorities. Other time to event endpoints like PFS, disease free survival (DFS) or event-free survival (EFS) have been accepted particularly when surrogacy to overall survival or clinical relevance can be demonstrated.[22]

For all time to event endpoints, the time between a triggering event (*e.g.* randomization or first dose) until a clinical event of interest occurs is analysed.

Typical time to event endpoints in oncology include:[11, 22]:

- **OS** is the time from randomization (or first dose, in non-randomized trials) until death from any cause
- **PFS** is the time from randomization (or first dose, in non-randomized trials) to objective disease progression, or death from any cause, whichever occurs first

- **DFS** is the time from randomization (or first dose, in non-randomized trials) to objective disease recurrence or death from any cause, whichever occurs first
- **EFS** is typically indication specific, but not necessarily consistent in different protocols. For example, for acute myeloid leukaemia EFS was defined as the time from randomization to (induction) treatment failure, relapse for those who have induction treatment success (*e.g.* complete remission), or death from any cause, whichever occurs first[23,24].

## 6. Missing data, intercurrent events, and loss to follow-up

During the follow-up of patients, situations might occur that either preclude the observation or affect the interpretation of the variable. These are defined in the ICH E9 addendum as intercurrent events. The guidance carefully differentiates missing data from intercurrent events:

> When most subjects are followed-up even after the respective intercurrent event (e.g. discontinuation of treatment), the remaining problem of missing data may be relatively minor. In contrast, when observation is terminated after an intercurrent event, which is obviously undesirable in respect of this strategy, the assumption that (unobserved) outcomes for discontinuing subjects are similar to the (observed) outcomes for those who remain on treatment will often be implausible.[1]

This distinction is made repeatedly in the ICH E9 (R1) guidance, which classifies the administrative closure of the trial as missing data. Its applicability to loss to follow-up in time-to-event contexts, however, could potentially be questioned. In the presence of non-proportional hazards, the end of the study can result in bias to estimators that assume proportional hazards, such as the hazard ratio, even absent intercurrent events.[20] The definition of intercurrent event in the ICH E9 (R1) glossary refers to events preventing the "existence" as well as the interpretation of measurements.[1] Loss to follow-up prevents the existence of future measurements.

Based in part on these considerations, Buhler, Cook, and Lawless proposed a terminology which classifies loss to follow-up as a kind of intercurrent event, a Type I intercurrent event, based on the definition of intercurrent event in the ICH E9 (R1) glossary.[25] While our paper retains the classification described in the body of ICH E9 (R1), certain pragmatic considerations in our approach, such as the handling of progression discussed in Section 8, tend to favor this alternative classification.

Even if not classified as an intercurrent event in its own right, loss to follow-up in clinical trials generally results from an event of some kind, such as the occurrence of progression or another event of interest, a patient decision to stop study assessments, the administrative end of the trial, a decision to put a patient on a new trial in the hopes of obtaining better efficacy, or a patient's failure to respond to further contacts.

All potential intercurrent events require an *a priori* assessment at the design stage to define the appropriate approach on how to deal with them along with a discussion how censoring is applied. An individual decision to discontinue treatment based on perceived lack of efficacy or toxicity, for example, will generally represent a potential intercurrent event.[1]

Both missing data and intercurrent events can occur while the patient is still being followed. For time-to-event variables, the observation period of a patient can be separated into the time up to the occurrence of an intercurrent event or endpoint of interest and the time thereafter. Just by the chronological sequence it can be concluded that the risk of the event of interest for the time interval up to the intercurrent event is not impacted by the intercurrent event but the time thereafter is. Classification of events that result in loss to follow-up as representing intercurrent events or missing data is accordingly particularly important under the estimands guidance.

In order to make this classification, it is critically important to collect necessary data, particularly data about the reasons for discontinuation or loss to follow-up for the relevant assessments. Under the estimands framework, cases where loss to follow-up has no apparent relationship to the treatment effect or underlying prognosis are considered as missing data. Cases where there is a potential relationship should generally be considered intercurrent events or integrated in other attributes of the estimand. Absent a clear scientific rationale for assuming otherwise, loss to follow-up should be considered as potentially related to treatment.

Given that clinical events may be informative, data collection on the outcome of interest and events that are not *a priori* defined intercurrent event (but may be considered as such with growing evidence) should continue where appropriate to the estimand and feasible.

## 7. Strategies for handling intercurrent events and some associated estimation methods

Both missing data and intercurrent events can occur without loss to follow-up. An isolated missed appointment does not generally prevent subsequent follow-up and documentation of the event of interest, is often unrelated to treatment effect, and is regarded as missing data. Many potentially biasing events – for example changes in relevant concomitant medications, – represent intercurrent events, but similarly do not result in a patient being lost to follow-up for the event of interest.

Frequently observed intercurrent events in oncology trials include deaths, discontinuation of treatment due to toxicity or subjective clinical progression and initiation of new anti-cancer therapy. When loss to follow-up occurs at the time of or due to intercurrent events, the assumption of non-informativity for resulting censoring is unlikely to hold.

Applicability of the ICH E9 (R1) Estimands guidance's strategies for addressing intercurrent events are described below. Treatment policy and hypothetical strategies generally impact the treatment definition, especially when the ICE is treatment related. Principal stratum strategies impact the population definition. Composite and while-on-treatment strategies impact the variable definition.

Different estimand strategies can be applied to intercurrent events, depending on the estimand of interest. It is important to keep in mind the estimands framework can involve more than one strategy per estimand. Within each estimand, different strategies can be and often will need to be applied to different intercurrent events.

**Treatment Policy Strategies**

The term "treatment policy strategy" comes from the idea of evaluating the effect of the complete sequence of treatments, beginning with randomization and including all treatments thereafter, through treatment switching[26]. This strategy, however, is not limited to treatment switching. It can potentially be applied to any intercurrent event. When applied to a general intercurrent event, the research question evaluates all outcomes up to the event of interest, through and beyond the applicable intercurrent event.

According to Lipkovich (2020), a treatment policy strategy "corresponds to the estimation of the total effect" of the originally assigned treatment, "where all post-randomization changes to treatment [we generalize to all effects of the applicable intercurrent event] are ignored and left unspecified" and regarded as effects of the original treatment assignment. In a randomized study, it "essentially assesses the effect of being randomized to treatment."[27] We would note that the strategy assumes and regards the estimator as targeting the total effect, whether or not this is actually the case. A treatment policy strategy considers all data observed relevant to the clinical question of interest. For reasons explained by Fleming et al.[28], when correctly executed, it is generally the strategy most likely to preserve the assumption that censoring is non-informative.

The treatment policy strategy is often preferred by regulators. The EMA's 2013 PFS/DFS Guideline Appendix[29] had stated that continuing to follow patients until the event of interest, through and beyond what were later to be called intercurrent events, was the approach most consistent with the original ICH E9 principle of intent-to-treat and constituted the default approach:

> Events of withdrawal from study therapy prior to adjudicated progression are likely to be informative and the adequacy of censoring these events in the statistical analysis should always be questioned. There is no way to handle this problem that is optimal for all situations, but the principles of intention-to-treat should be followed as far as possible when defining the analysis set for the primary analysis of PFS/DFS. In particular, for all randomised patients, outcome data should be collected according to the intended schedule of assessment and the date of progression or recurrence should be assigned based on the time of the first evidence of objective progression or recurrence *regardless of violations, discontinuation of study drug or change of therapy*.

> If, for a particular study, a different approach is considered to be more
> appropriate, a justification is expected and CHMP Scientific Advice
> agreement is recommended at the planning stage.[29] [emphasis added]

However, the strategy is not without limitations. Because per the ICH E9 (R1) addendum, a treatment policy strategy considers the applicable intercurrent event irrelevant to the outcome and looks to what occurred after it, valid execution of a treatment policy strategy requires follow-up consistent with this approach.

As the ICH E9 Guidance explains, a terminal event generally defeats a treatment policy strategy:

> In general, the treatment policy strategy cannot be implemented for
> intercurrent events that are terminal events, since values for the variable after
> the intercurrent event do not exist. For example, an estimand based on this
> strategy cannot be constructed with respect to a variable that cannot be
> measured due to death.[1]

However, the set of events that defeat a treatment policy strategy is not limited to terminal events. Systematic and pervasive failure to follow patients through and beyond intercurrent events that introduce or reflect a change in hazards can also result in a biased estimate of a treatment policy estimand. Whether or not a process termination occurs, attempting to infer the entire distribution from the portion sampled can result in a biased estimate.

Accordingly, execution of a treatment policy strategy requires consistent follow-up through and beyond intercurrent events until the event of interest is observed. This degree of follow-up can be challenging to obtain in some clinical trials and may be infeasible in some cases. For example, patients may wish to withdraw from a trial following treatment discontinuation, or for reasons related to treatment efficacy or safety. Thus, informative censoring is likely unavoidable.

A distinction should be made between situations where failure to follow patients becomes pervasive and systematic, and situations where it remains relatively isolated and occasional. Occasional informative censoring is practically inevitable in clinical trials and does not change the interpretation of the underlying implied estimand. However, where failure to follow patients is pervasive and systematic, the assumptions required for valid execution of a treatment policy strategy may be infeasible, and a strategy that does not require or assume such follow-up will occur in most patients may be more appropriate. As we explained in the terminology section, pervasive and systematic failure to follow patients beyond intercurrent events effects an artificial process termination impacting the implied estimand, and tending to defeat the validity of the estimator for the original estimand, if a treatment policy strategy was employed.

There has been a line of research on models that impute effects for patients not followed beyond intercurrent events based on similar patients who were.[30,31,32] Unlike hypothetical-strategy models discussed below, these models attempt to estimate effects through and beyond the intercurrent event based on similar patients who were actually followed through and beyond it, and hence can be classified as within the realm of treatment policy strategy estimation. These models may help reduce the proportion of non-followed patients while still permitting estimation of a total treatment effect. Further improvement might reduce the needed proportion further. However, it remains the case that a substantial portion of patients must continue to be followed past the intercurrent event for these models to work, and hence systematic failure to follow defeats a treatment policy strategy. In addition, the assumptions required for these models to be valid may themselves be questionable.

Where it is infeasible to follow patients through and beyond intercurrent events that introduce informative censoring, and this loss to follow-up is expected to be pervasive and systematic, an alternative to the treatment policy strategy should be considered.

**Composite Strategies**

Composite strategies make intercurrent events part of the variable (or endpoint), e.g. by making their presence equivalent to a "bad" outcome on the scale of interest. In the context of time-to-event estimands, a composite strategy makes the event of interest a composite event that includes both the original event and the intercurrent one. This strategy is particularly useful when the intercurrent event is positively correlated with, and highly informative of, the event of interest. Perhaps the best-known example in oncology is PFS, a composition of progression and death events. Composite strategies may be appropriate in a wide variety of contexts. Composite strategies handle intercurrent events in a manner that is generally both statistically valid and operationally feasible, and is often (although not always) clinically meaningful. Making the intercurrent event a component of the event of interest eliminates validity problems that might arise due to loss to follow-up and associated informative censoring as a result of the intercurrent event. In addition, a composite strategy tends to increase the likelihood of events and reduce time to event, increasing the reliability and power of estimation and decreasing study duration. Accordingly, a composite strategy should be considered when continuing to follow patients beyond the intercurrent event may be infeasible.

**Hypothetical Strategies**

Hypothetical strategies ask what would have happened in a counterfactual scenario, *e.g.*, if the intercurrent event had not occurred. Such a strategy may be particularly appropriate when the relevant intercurrent event only occurs because of experimental conditions and would not occur in real-world clinical practice.

Hypothetical strategies tend to remove events from the clinical question of interest. Generally speaking, events following the intercurrent events are ignored.

Our first and simplest example of implementing a hypothetical strategy is a familiar traditional practice in the pharmaceutical industry, explicitly "censoring" for intercurrent events. The 2007 FDA Cancer Endpoint Guidance mandated censoring for subsequent therapy.[10] This implementation assumes that future hazards of patients with the intercurrent event are the same as patients without it.

We think the underlying assumption is generally unreasonable in an oncology setting, and we recommend reconsidering this practice. In the context of treatment switching, the traditional practice asks what the treatment effect would be if change of therapy had not occurred, assuming that patients with new therapy have the same future hazard of experiencing an event as those without it. However, the reason patients need a change in therapy is generally that they are perceived as less likely to be benefiting from the treatment, and different therapy will likely result in different efficacy. The practice often results in pervasive informative censoring.

As discussed below, it has been argued that this traditional censoring practice can be interpreted as implementing an implied independent-causes while-on treatment strategy.[9] However, absent specific reasons to think the non-informativity assumption plausible, assuming a scenario where it does hold effectively assumes a hypothetical scenario. For this reason, we recommend considering alternative approaches. If such an estimator must be interpreted post-hoc, a hypothetical implied estimand is the better interpretation.

More sophisticated hypothetical approaches based on counterfactuals, such as inverse probability censoring weighting (IPCW) or rank preserving structural failure time (RPSFT), are also available. Counterfactual survival times are survival times that would have occurred if the intercurrent event had not happened, as estimated by a model.[7] Considerations for these methods, and contexts when use of a hypothetical strategy might be appropriate to consider, are discussed in the next section.

**While-on-Treatment Strategies**

A while-on-treatment strategy is only concerned with what happens up to the time of the intercurrent event. In a while-on-treatment strategy, anything after the intercurrent event is considered irrelevant to the clinical question of interest, and hence data collection need not continue past it. The strategy can be applied to any intercurrent event, not just withdrawal from study treatment. As ICH E9 (R1) notes, "Terminology for this strategy will depend on the intercurrent event of interest; e.g. 'while alive,' when considering death as an intercurrent event."[1] The same would be true of other types of intercurrent events, e.g. "prior to surgery," "prior to progression," etc. For this reason, it might be better to name this strategy "while prior to intercurrent event".

A motivating example is palliative treatment, which is generally not assumed to prolong survival. Clinical interest lies in evaluating alleviation of symptoms prior to death, whenever death occurs. However, a while-on-treatment strategy may be appropriate for many other contexts where what happens after the intercurrent event is not considered relevant to the research question.

We briefly discuss implementation of a while-on-treatment strategy in the context of time-to-event estimands. Implementation needs to evaluate whether the intercurrent event can be assumed independent of the underlying treatment effect. When independence can be assumed, then a while-on-treatment strategy can be implemented through a cause-specific hazards model, which uses standard censoring and hence represents an interpretation of standard proportional hazards and related censoring models and methods (e.g. Kaplan-Meier, Cox, log-rank).[9] In this case the implementation censors for the intercurrent event, and does so without quotation marks. As discussed above in the context of hypothetical strategies, causal independence can rarely be reliably established in oncology, and time to cause-specific event estimands should be used only when such independence is plausible. Independent causes might for example be appropriate for a safe drug for early-stage cancer, where any deaths could be assumed not caused by the underlying cancer or treatment, or perhaps for time to cause-specific death where causality attribution can be reliably established and independence is a reasonable assumption,

When the independence assumption lacks a clearly plausible basis, which we suggest is often and indeed usually the case in oncology, the situation is more complicated. Unkel (2019) in the context of a narrow view of "while on treatment," suggests that a while-on-treatment strategy can be implemented using a competing risk method with the intercurrent events of treatment discontinuation or death classified as competing events.[9] Generalizing this, the approach could be applied to any intercurrent event. A dependent-causes competing risk model is generally implemented through methods like the Cumulative Incidence Function (CIF) and the subdistribution hazards model.[33,34]

There is no general agreement that a dependent-causes competing risk model actually implements a while on treatment strategy. It does not implement a treatment policy strategy, as it readily fits terminal events like death for which treatment policy strategy cannot be implemented.[1] Perhaps dependent-causes competing risks models could be regarded as implementing an entirely separate strategy. We suggest that classifying as a while-on-treatment strategy represents the best fit for the Guidance's current list. This type of competing risk model reflects not just the strengths but the weaknesses of while-on-treatment strategies generally, as described in other contexts. A CIF represents the cumulative incidence of the event of interest up to the point of the intercurrent event; with no concern for what happens after. As the Guidance states, "particular care is required if the occurrence of the intercurrent event differs between the treatments being compared"[1] Hahn and Zhou (2023) use a palliative-therapy example to illustrate while-on-treatment strategies in a longitudinal context. As they explain, "if the pain for patients with chronic diseases increases with time, a poisonous drug that can kill people in a relatively short time could produce better results than a placebo, which is misleading."[35] In a time-to-event context, dependent-causes competing risks methods have exactly this weakness that Hahn and Zhou found characteristic of while-on-treatment strategies generally. Patients can appear to have a lower cumulative incidence of pain or symptom progression simply by dying sooner. A clear understanding of the medical and ethical meaning and appropriateness of such a strategy is accordingly particularly important before using such a strategy.

One practical way to address the susceptibility of while-on-treatment analysis to confounding a greater hazard of the intercurrent event with a lower incidence of the event of interest would be to first analyze an estimand that has the intercurrent event as its event of interest. For example, if survival benefit or non-inferiority is first established, an analysis of time to pain progression treating death as a competing risk could then be done with some protection from risk that any apparent benefit is due to patients dying early.

The subdistribution-hazards (Fine-Gray) model sometimes used for modelling and testing in this context has been criticized as not having a causal interpretation. However, the descriptive Cumulative Incidence Function (CIF) does not have this issue.[33]

**Principal Stratum Strategies**

A principal stratum strategy[36,37] attempts to study only a population in which the intercurrent event of interest would not occur regardless of assigned treatment. It is generally implemented by using a model to predict such patients from baseline characteristics and removing patients expected to experience it from the study sample. Once the set of patients in whom the applicable intercurrent event is expected not to occur is identified by the estimating model, ordinary censoring can be applied to patients for whom the intercurrent event occurs despite the model's prediction that it will not. Just as too high a proportion of patients no longer being followed due to an intercurrent event can invalidate a treatment policy strategy, too high a proportion of patients experiencing the intercurrent event, while not changing the strategy, can render the model implementing a principal strategy nonpredictive and invalid. A principal stratum strategy has not been common in regulatory oncology clinical trials.

8.         Kinds of intercurrent events

*<u>Progression.</u>* Our paper discusses progression of various kinds as if it were an intercurrent event for other events of interest. Progression itself generally reflects a worsening of the patient's condition with a poorer prognosis, and hence tends to indicate a likely change in hazards for other efficacy-related events such as death. However, it generally reflects an anticipated change in the underlying disease and treatment effect dynamics rather than an unexpected event or intervention and, accordingly, is not necessarily itself an intercurrent event. We have included it in our discussion of intercurrent events as a pragmatic classification to aid study design, because, as Unkel noted,[9] progression often results in systematic withdrawal from clinic visits and induces systematic informative censoring. It accordingly often affects the implied estimand and tends to require handling similar to intercurrent events for other oncology estimands. It will accordingly generally need to be treated in the same manner as intercurrent events when designing clinical trials. We would note that this pragmatic rationale for handling progression as an intercurrent event tends to support the practical utility of Buhler, Lawless, and Cook's classification of loss to follow-up in general as a type of intercurrent event.[25]

When other estimands depend on clinic visits that end at progression, we recommend an explicit strategy for addressing progression's informative effects rather than simply relying on censoring at the end of assessments. Where clinically appropriate, a *composite strategy* might be the simplest approach in terms of clear statistical validity. An explicit *hypothetical strategy* could be considered, although the assumptions underlying implementing such a strategy through traditional "censoring" are unlikely to be met. A *while-on-treatment strategy* implemented by treating progression as a dependent-causes competing risk might be considered, if its weakness that lower cumulative incidence of the event of interest may result from earlier progression does not render it inappropriate under the circumstances. Approaches and recommendations for handling of typical kinds of intercurrent events within oncology trials are as follows:

<u>*Death*</u>. Death is the quintessential terminal event. Accordingly, a **treatment policy strategy** cannot be used for death, and another strategy should be used. The interpretation that systematic censoring for intercurrent events induces an implied **hypothetical strategy** does not apply here, as the research question underlying a hypothetical strategy, what would have happened if the patient had not died, is not generally a meaningful clinical question. While censoring for death could potentially be interpreted as an implied **while-on-treatment (while-alive) strategy** implemented assuming independent causes, this strategy would only be valid if death is truly independent of the event of interest, which in unlikely in the context of a disease with a high mortality outcome. On the other hand, a **composite strategy** is often a good choice due to the tendency for death to be highly correlated with other clinical events. Perhaps the most common example of a composite strategy involves time to progression, which has largely been replaced by the composite endpoint progression-free survival. In situations where there is only concern about the event of interest prior to death, or when survival superiority or non-inferiority has been previously established, (e.g. a secondary estimand in a trial with a primary survival estimand), a **while-on-treatment (while alive) strategy** could be implemented without assuming non-informativity or causal independence, by using a competing risk approach as estimator. An example would be the competing risk approach to time to bone fracture with death as a competing risk event.[36]

<u>*Discontinuation of treatment due to toxicity*</u>. Participants are increasingly being followed beyond treatment withdrawal, rendering a **treatment policy strategy** generally feasible. For some endpoints of interest, a **composite strategy** may be appropriate. Where patients are not followed past end of treatment, traditional explicit censoring at treatment discontinuation, explicit or implicit, would induce an implied **hypothetical strategy**. The research question involved, what time to the event of interest if patients had not experienced toxicity and not withdrawn from treatment, could in some cases be relevant. However, implementing such a strategy through simple censoring would generally not be appropriate, as it is often not reasonable to assume that patients with early discontinuation from treatment would have the same future prognosis as patients not discontinued. In cases where it is inappropriate or infeasible to follow patients beyond treatment withdrawal, either the real clinical question of interest may not be concerned with what happens afterwards, or not being concerned with what happens afterwards may represent a compromise clinical question that can feasibly be addressed under the circumstances. In this case a **while on treatment strategy**, implemented using a dependent-causes competing risks approach as estimator, might be appropriate.

*Subjective clinical progression*. Oncology clinical trials are increasing requiring follow-up until documented radiological progression as clinical progression has been criticized as an ill-defined, potentially subjective, and potentially biased assessment.[39] Many trials offer the option for participants meeting pre-specified criteria who are in the opinion of the investigator continuing to derive benefit to remain on treatment after the initial documentation of radiographical progression. In such cases, where it is broadly feasible to follow participants until documented progression, a **treatment policy** strategy is appropriate. However a **composite strategy** might be considered in cases where clinical progression prior to radiological progression is expected to occur frequently and it is not practical to continue to follow participants, particularly where clinical progression is highly predictive of radiological progression or death. In some cases, potential for bias due to the subjective nature of clinical progression might need to be balanced against bias due to systemic inability to follow patients resulting in informative censoring.

*Central vs. local review.* In oncology trials, it is not uncommon to have a retrospective central radiological review in addition to a local review by the investigator. One of these reviews, generally the central review, is considered primary. There is often a discrepancy between central and local review results. When central review is retrospective, patient management decisions including decisions to discontinue tumor assessments are based on local review. When a patient progresses per local review, further imaging is not available to the retrospective central review, resulting in PFS being censored if the central review finds no progression. But when central review finds progression before the local review, both reviews are complete. Fleischer (2010) discussed the issue and hypothesized a model under which this difference introduces bias into the results.[40] Zhang (2013), however, evaluated a meta-analyses of studies and concluded that retrospective central review introduces variation but not bias.[41]. We would note that both local and central review are addressing the same clinical question. The introduction of real-time central review, in which the central review decision is conducted rapidly and the results are considered by the investigator in making treatment decisions, can alleviate the issue, although the one-way transmission of information that occurs in the process can result in the central review results informing or replacing local review. Central audits, where only a portion of patients are sampled, represent an alternative intermediate approach where real-time central review is not feasible. The secondary review (generally the investigator review) should be considered a sensitivity analysis.

*Incorrect medication*. While it is not uncommon for patients not to be able to tolerate or otherwise depart from the planned study dose, receipt of the wrong study medication in a trial is generally a rare, isolated event. As such, even though patients with incorrect treatment are often discontinued from most study assessments, and resulting censoring may well be informative, these events are not likely to be pervasive or systematic enough to affect the overall interpretation of the results or choice of estimand. Accordingly, a **treatment policy strategy** can generally be used. In the unlikely event that incorrect medication is a pervasive issue in a study, perhaps a supplementary **hypothetical strategy**, implemented using a causal inference method, might be considered to help assess what would have happened if incorrect medication had not been given, However, such an outcome would tend to suggest poor study execution. Results of a poorly executed study may be unreliable regardless of how evaluated.

*<u>Initiation of new anti-cancer therapy.</u>* When the subsequent therapy given in a trial is of a sort that would be given in normal clinical practice if the product was approved, and if patients can generally be followed beyond subsequent therapy to the event of interest, a **treatment policy strategy** can be used.

However, an increasing problem in contemporary trials is a situation where the trial conditions induce behavior that would not otherwise exist in the clinic. Examples discussed in Manitz et al (2022)[8] include:

- In blinded trials, when there are multiple experimental treatments being studied in the same class, patients assigned to the control arm may later enter another trial evaluating an experimental treatment in the same class.
- In open-label trials, patients not assigned to the desired treatment may immediately withdraw without receiving study treatment, or otherwise withdraw earlier than would be the case in regular clinical practice. For example, in Checkmate 037, 23% of patients randomised to the control arm did not receive assigned treatment compared with 1% randomised to active therapy[9].

Both of these situations would not normally occur in the clinic. Patients who have progressed on a therapy would not normally take a therapy in the same class as subsequent therapy. And patients who begin a treatment regimen for a serious cancer would not normally immediately switch without cause. These situations are therefore artifacts of trial conditions, based on the fact that in a trial, patients are randomly assigned a treatment which may not be the one they were hoping to receive, but nonetheless have the ability to "vote with their feet" if they do not like the assignment. They thus violate an implicit assumption of a treatment policy strategy, that the treatment pattern observed in the trial predicts the treatment pattern that will occur in the clinic in the event of approval.

Where the context of a clinical trial results in conditions and induces patient choices not likely to be repeatable in real-world post-approval clinical practice, a **hypothetical strategy** addressing the counterfactual scenario in which such patient choices did not occur may better reflect the clinical question feasibly addressable by such a trial. Such a hypothetical strategy might be considered, implemented with causal inference methods such as rank-preserving structural failure time (RPSFT) to estimate the outcome had patients not crossed over from control to active treatment;[42] the 2-stage method to estimate the outcome had patients not crossed over at a specific disease-related time point such as progression;[43] and inverse probability censoring weighting (IPCW) to estimate the outcome in the absence of new therapy.[42]

These methods have been subject to criticism because of the strength of the assumptions required, particularly the assumption of no unmeasured confounding. While the reliability of the answer provided by hypothetical strategies may be criticized and indeed cannot be assured, they have the advantage of addressing a clinically relevant question. See Manitz (2022) for a more detailed discussion.[8]

As ICH E9 R1 notes, "usually an iterative process will be necessary to reach an estimand that is of clinical relevance for decision making, and for which a reliable estimate can be made."[1] Subsequent therapy may represent an example of a conflict between clinical relevance and reliable estimation, requiring care in selecting a clinical question and intercurrent event strategy representing a reasonable balance between the two.

***Surgery and stem cell transplant***. Surgery may occur during oncology trials for multiple reasons. For example, surgery is a planned procedure in neoadjuvant trials, or as palliative or curative treatment in later stage disease. For an EFS endpoint, surgery could be a component of a **composite strategy**. Alternatively, if unplanned surgery is potentially curative, a **treatment policy** approach may be considered, following the patient beyond surgery until the event of interest. In the case of palliative surgery, this could be handled similarly to initiation of anti-cancer therapy described above. Similar considerations apply to stem cell transplant as an intervention.[44] It may not always be possible to follow patients beyond surgery or stem cell transplantation. The traditional practice of "censoring" at the time of the procedure would induce an implied hypothetical strategy, and would have issues similar to traditional "censoring" for other subsequent therapy.

In addition to these typical intercurrent events which occur in oncology clinical trials, there might be intercurrent events observed during the course of a trial which could not be foreseen at the design stage and which cannot be controlled by study procedures, like the occurrence of a pandemic.[45] Such cases require careful re-assessment of the estimand definition.[1] In this case, as in others where there is a possibility that patients might be less able to be followed than expected or hoped for at the beginning of the trial, supplementary estimands addressing this possibility should be considered.

## 9. Practical consequences for study design and data collection

In the estimands framework, data collection is as important as design, as the estimand definition(s) in a trial determine how long rigorous data collection is required. Additionally the chosen estimator will generally require data on intercurrent events. This may require augmentation or even redesign of standard data collection systems and procedures. Loss to follow-up, whether or not itself an intercurrent event,[25] is often caused by intercurrent events. It can also cause intercurrent events, since patients ending study participation may change or end treatment or follow-up care. For these reasons, special attention is needed in the study design to forecast the likely potential intercurrent events that will result in loss to follow-up The CRF should capture relevant data documenting loss to follow-up, including identifying intercurrent events and their likely relationships to treatment effect that might be associated with loss to follow-up.

Data collection for each assessment should include whether the assessment occurred, reason for no assessment, and a causality (relatedness to study treatment). This causality assessment could be similar to the one commonly used to determine treatment relatedness of AEs. The list of reasons for no assessment should not be limited to predefined intercurrent events. The underlying reason should be captured in sufficient granularity to identify intercurrent events not being considered relevant at the design stage.

This requirement is particularly applicable to documenting withdrawal from study treatment or change in therapy, but also to withdrawal from assessments underlying key estimands. For example, if tumor assessments, biomarker assessments, and safety labs continue following treatment but end at different times, we recommend collecting information on each instance separately. We believe that better documenting reasons for withdrawal from study treatment, clinic visits, and particular assessments represents an opportunity for improvement in current practise. Currently, reasons for withdrawal from treatment and the like are often singularly not informative in assessing treatment relatedness, with reasons such as "patient withdrew consent," "investigator decision," and similar common. We recommend categories that clearly indicate, wherever ascertainable, the investigator's opinion as to whether reasons for withdrawal were or were not related to perceived treatment efficacy or safety.

Investigator opinion on causality can be subjective and unreliable. Recognizing collecting subjective causality information is imperfect, better collection with more relevant categories nonetheless represents an improvement over current conditions. With attention to this issue, further improvements still are likely possible.

ICH E9 (R1) states: "A prospective plan to collect informative reasons for why data intended for collection are missing may help to distinguish the occurrence of intercurrent events from missing data. This in turn may improve the analysis and may also lead to a more appropriate choice of sensitivity analysis."[1] Implementing this prospective plan, documenting the existence and reasons for missed assessments, together with a causality assessment (relatedness to treatment), represents a significant change from current practice. However, collecting this additional data to support the needed sensitivity analyses is important in the context of the estimand framework. This recommendation may be particularly relevant to time-to-event estimands that are the basis of label claims.

Alternative approaches to data collection might be considered to increase follow-up beyond certain types of intercurrent events. For example, remote visits and digital devices can collect physical function and patient diary data to replace frequent clinic visits to complete questionnaires. Such automated and at-home data collection avoids additional burden for patients and minimizes the risk to stop follow-up prematurely. Improvements in technology have already aided data collection outside clinic visits and are likely to continue to improve.

**10.         Addressing when patients cannot be followed: The importance of design**

Traditionally, oncology clinical trials ended clinic visits at events like treatment discontinuation or progression, with only long-term follow-up (typically telephone contacts) afterwards. In addition to the estimand framework, an evolving interest in understanding the impact of treatment on how patients feel and function as well as an evolving therapeutic landscape often requires additional data to understand whether or not the natural history of the disease is worsened by the study drug and to demonstrate if there can be benefit beyond initial disease progression. These concerns have resulted in an increased interest in following patients past intercurrent events. However, it is recognized that there are often practical limitations. Additionally, it is recognized that required follow-up within a clinical trial may focus on the primary or key secondary endpoints and in an effort to reduce burden on trial participants limited considerations may be given to follow-up for other supportive endpoints. For example, in many oncology trials follow-up post progression may be substantially reduced to focus on remote assessments to collect survival and subsequent treatment. In such cases continued follow-up on patient reported outcomes may be halted limiting the ability to understand the overall impacts of study treatment.

Wherever possible, study design teams should consider alternatives to clinic visits, such as electronic diaries, telephone or electronic communications methods, and/or home visits, wherever continued follow-up is indicated by the desired strategy but there is reason to suspect that continuing clinic visits will lead to systematic loss to follow-up. New technology and methods permit an increasing range of alternatives. It is possible that in the future the problem of inability to follow patients will disappear entirely. However, that future is not yet here. Oncology trials are still replete with assessment methods that cannot be done at home. For the time being, the end of clinic visits means the end of assessments for a number of standard variables and indicators.

The ICH E9(R1) guidance provides a useful framework for addressing cases where it may not be practical or feasible to continue follow-up. It provides multiple strategies and options for addressing intercurrent events, both options that require following patients beyond them, and options that do not. These options may provide a navigable path to defining a clinical question of interest that can be addressed in the context of the disease setting.

An additional practical issue in study design involves subordination of follow-up for lower-priority study objectives to the needs of higher-priority ones. It is not uncommon in oncology clinical trials for data collection considerations based on a primary variable to affect data collection for secondary variables. For example, if the primary objective for a particular study involves PFS, the assessment schedule will typically be based on this objective. As Unkel observed,[9] clinic visits, even when extending past subsequent therapy, may need to end at documented radiological progression. Ending clinic assessments at progression will also end data collection for secondary estimands requiring in-clinic assessments, for example time to forced lung capacity deterioration and time to symptom improvement or deterioration. This can effect a systematic artificial process termination, preventing reliable estimation of a treatment policy strategy.

Where assessment ends at an event based on the needs of a different, higher-priority study objective, and the ending event may be informative with respect to the event of interest, we recommend that the study design team explicitly acknowledge the event resulting in loss to further clinic visits as an intercurrent event, and devise an appropriate strategy that is both scientifically reasonable and feasible in the context. An explicit hypothetical strategy would be preferable to leaving the effect of the assessment-ending event unacknowledged. But other strategies acknowledging the impact of the event on interpretation may be more appropriate, particularly where the relevant event is highly informative. It might, for example, be appropriate to consider a composite strategy and assess time to the earlier of symptom deterioration or progression, whichever occurs first. In some cases, the lowered priority of the estimand may reflect the fact that what happens after the assessment-ending event is not of sufficient interest to be worth study in the context of the particular clinical trial, and a while on treatment (i.e. while prior to occluding event) strategy may be appropriate.

Compromises are inevitable in applied clinical research. It is important for the design team to first identify the clinical question of interest and understand the optimum strategy to address that question. Once that is done, the team should proceed to investigate whether the clinical conditions, estimand priority within the study, and other factors render the strategy feasible in the context. If a compromise is necessary, the team should be conscious of both what is desired and what is possible. It should understand the limitations compromise imposes on the ability of the study to address the original question, including understanding how changing the strategy changes the research question, and how design features can implicitly change the strategy.

The Estimands Guidance notes that "usually an iterative process will be necessary to reach an estimand that is of clinical relevance for decision making, and for which a reliable estimate can be made." Figure 1 illustrates a proposal for an iterative approach to construct an estimand that is both of clinical relevance for decision making and operationally feasible at the design stage. This iterative approach is similar to Deming's Plan-Do-Study-Act cycle.[46]

**Figure 1**: Illustration of an iterative approach to construct an estimand that is of clinical relevance for decision making and operationally feasible.

[Figure 1 here]

## 11. Sensitivity and supplementary analyses

ICH E9 (R1) defines sensitivity analyses as "a series of analyses conducted with the intent to explore the robustness of inferences from the main estimator to deviations from its underlying modeling assumptions and limitations in the data."[1] The purpose of a sensitivity analysis is to check the assumptions underlying the estimand. This concept of a sensitivity analysis represents a change in meaning from common past use. The guidance requires deviation from assumptions to be checked for "comprehensively."[1]

Every strategy for addressing intercurrent events is based on assumptions and requires conditions to be valid. The purpose of sensitivity analyses is to check, to the extent feasible, whether the applicable assumptions required for the primary estimand are reasonable under the circumstances. As ICH E9 (R1) states, "Inferences based on a particular estimand should be robust to limitations in the data and deviations from the assumptions used in the statistical model for the main estimator. This robustness is evaluated through a sensitivity analysis."[1]

We recommend that for each estimand, study teams should identify key assumptions made in the selection of the estimand and associated estimator and identify an associated sensitivity analysis checking for it. Sensitivity analyses can include alternative analyses for the same estimand using different assumptions. They can also include "model-checking" type analyses that check specific assumptions directly.

The discussion here focuses on sensitivity analyses, as defined in ICE E9 addendum, related to survival-related assumptions. Sensitivity analyses are a particular problem in a survival context.

A key element of sensitivity analyses in a survival context is to check the appropriateness of non-informative censoring assumptions, and hence whether presumed "missing data" can in fact legitimately be so characterized. Events that occur after censoring are not always documented and it is not possible to know whether censoring is informative. Similarly, assumptions needed for a hypothetical analysis are often unverifiable. Nonetheless standard censoring checks and sensitivity analyses are possible, and commonly employed.

Examples include:

- When using a model, sensitivity models with alternative parametrization or parametric assumptions. (e.g a Weibull model or a piece-wise exponential model as a sensitivity analysis for a Cox model).

- Tipping point analysis to assess censored-at-random assumptions (e.g. Atkinson et al., 2019)[47]

- Descriptively assessing the distribution of censoring to see if it occurs evenly between the arms.

- Analyses with and without censoring of events that are observed following more than one missed assessment. While this analysis was traditionally performed regardless of reason for missed assessment, it might be appropriate to perform an additional analysis focusing on assessments missed due to possible intercurrent events (per CRF data collected as recommended in Section 9).

- Checks for key traditional statistical assumptions such as proportional hazards.

- Interval censored analyses (potentially a different estimand and hence a supplementary analysis, but also can be used to check the appropriateness of right-censoring).

As discussed above, a common issue in survival trials is the use of treatment policy strategies where systematically following patients beyond the intercurrent event of concern may not be feasible. We recommend sensitivity analyses to address this assumption. This could be done with simple descriptive methods. For example, for PFS studies with a treatment policy strategy for events like end of treatment, clinical progression, or change of therapy without radiological progression, we recommend identifying what proportion of patients had these events prior to progression, and what proportion did not receive further tumor assessments beyond these events. Large differences in the number of events between arms could lead to concern. Sensitivity analyses could assess different assumptions for patients who were lost to follow-up due to other, e.g. unambiguously terminal events. For example, patients who died shortly after change of therapy could be assumed to have received full follow-up regardless of the change.

Supplementary analyses, by contrast, are less clearly defined by the estimands guidance. ICH E9 (R1) states that "distinct from sensitivity analysis, where investigations are conducted with the intent of exploring robustness of departures from assumptions, other analyses that are conducted in order to more fully investigate and understand the trial data can be termed 'supplementary analysis.'".[1] Lynggard (2022) states that "ICH E9(R1) does not clearly state which estimand a supplementary analysis targets and there is currently no consensus."[48]

We suggest that, to distinguish supplementary from sensitivity analyses, it would be useful for practitioners to regard supplementary analyses as targeting a different estimand in order to provide additional insights into the treatment effect. They may evaluate e.g. a different population, a different definition of the underlying event or variable of interest, or a different strategy for addressing intercurrent events. For example, these could include subgroup analyses, exploration of the components of a composite endpoint, or various alternative definitions of progression (e.g. including clinical progression). They could also include analyses using e.g. a hypothetical or while-on-treatment strategy when the primary analysis was based on a treatment policy strategy.

With this proposed clarification, many of the analyses traditionally labelled "sensitivity analyses" become kinds of supplementary analyses under the estimands framework.

Our paper has focused on cases where patients cannot be followed as planned at the beginning of the trial. We recommend supplementary analyses covering this possibility as a general practice in trials with time-to-event estimands.

**12.     Discussion and conclusions**

In the past, the assumption of non-informative censoring has rarely been challenged by regulatory authorities. Except in special cases like the pre-planned maturation of the trial, censoring is often informative. Its traditional widespread use to estimate in the presence of intercurrent events has often ignored the potential to bias results. The estimand framework addresses this issue. by focusing closely on the research question to be addressed, ensuring estimators, study design and visit schedules are appropriate to the question. The framework's alternative strategies for identifying and addressing intercurrent events provide methods for handling situations where implementing a traditional "ITT" or a treatment policy approach would lead to pervasive informative censoring.

Different approaches are necessary to address different clinical questions of interest. As discussed in Sections 9 and 10, the approach to study design should begin with the clinical question of greatest interest, but may require an iterative process to ensure alignment of the clinical question of interest with the resulting estimand, and to ensure that both can be validly supported by data that can be feasibly collected in the planned trial.

In general, the treatment policy strategy, regardless of intercurrent events, has become the default standard and has generally been recommended by regulatory authorities for randomised pivotal clinical trials where data can be consistently and systematically collected until the event of interest or study termination. From an estimands perspective, the treatment policy strategy reflects the entire treatment regimen, including subsequent therapies. Other strategies such as composite, hypothetical, or while-on-treatment strategy might be considered when the question of interest differs or where the assumptions underlying the treatment policy are not met. A treatment policy strategy will not insulate the trial from confounding, and events such as subsequent therapy that are inconsistent with real-world practice might better be handled as confounding intercurrent events than treated as censorable. It is important to recognize that changing the strategy and the handling of intercurrent events changes the estimand and its interpretation.

When the treatment policy strategy is not used, explicit "censoring" tables traditionally common in the pharmaceutical industry are generally replaced by another mechanism for handling intercurrent events at the estimand rather than the estimator level. For example, composite strategies may handle intercurrent events as a component of the event of interest; while-on-treatment strategies may handle them as a competing risk event; hypothetical strategies may be implemented with a causal inference model; and so on.

Our paper emphasizes the problem of process termination, pervasive or systematic patterns of loss to follow-up that defeat a treatment policy strategy, but which are not accounted for in the study design or estimand specification. In the authors' experience, clinical trial protocols are often replete with withdrawal criteria, censoring rules, and other features which result in systematic loss to follow-up and artificial process termination, without taking the effect of these features on the meaning and interpretation of the results into account. Eliminating this tendency, and ensuring that cases where patients withdraw or are lost to follow-up in a manner that can affect the interpretation of the results, can represent an important potential practical benefit of the estimands framework in improving trial design and conduct. Accordingly, this represents an important weakness in existing clinical trial practice that could be improved by a comprehensive application of the estimands framework that takes all features of study design and conduct into account. A study may be described as using a treatment policy strategy, but in fact on closer examination the study design shows that the protocol or patient-management practice results in patients being systematically withdrawn for some relevant intercurrent event, often radiological or clinical progression.

We stress that a treatment policy strategy does not consist of merely applying the label and refraining from applying traditional explicit censoring rules for subsequent therapy or other intercurrent events. Rather, it requires that patients actually can be and are followed, systematically, to the event of interest.

Trialists should endeavor to reduce, where possible eliminate, and appropriately account for the occurrence of artificial process termination in registrational clinical trials by carefully evaluating the implications of withdrawal criteria, assessment schedules, and anticipated patient practice on the results, including loss to follow-up and censoring patterns that are likely to arise as a consequence. Events which terminate assessments or trigger withdrawal criteria, such as progression in many studies, are particular candidates. When the study design systematically stops assessments at a particular event, we recommend explicitly identifying that event and determining an appropriate strategy to be used for that event. Applying catch-all censoring at end of assessments in situations where patients are being pervasively and systematically removed from follow-up by design or widespread practice, without investigating effect on research question or interpretation, will not result in a valid estimate. A key purpose of the estimands framework is to ensure, prior to the beginning of the trial, that estimation will be valid to the extent feasible.

It is accordingly necessary to ensure alignment of study design with goals and the requirements of the corresponding estimand up-front. Alignment requires a dialogue among clinicians, statisticians, and trial managers on the design team to clarify the research question and discuss how proposed withdrawal criteria, visit schedules, and anticipated patient and investigator behaviors could affect the interpretation of the study and limit the possible estimands that could be applied. This conversation should take place early, before study design proceeds very far.

The purpose of a clinical trial is often to predict real-world clinical practice, especially for a registrational trial. Certain elements of a clinical trial, such as randomization, are not reflective of real-world practice, and may induce patent behaviour not replicable in the real world. Where this occurs, a treatment policy strategy might not be the most appropriate. A hypothetical strategy addressing the counterfactual scenario of what would have happened if the non-real-world behaviour had not occurred, might be more relevant to real-world practice and hence might sometimes be preferable, despite problems with establishing the reliability of causal inference methods.

Rigorous data collection and trial monitoring are the key to addressing and distinguishing missing data and intercurrent events. The estimands framework depends on good data collection, including data about missed assessments. Events that are not collected cannot be managed or addressed. It is therefore critical to obtain the reasons for and dates of withdrawals and losses to follow-up, and to obtain data permitting assessment of the existence and dates of underlying intercurrent events.

Under both classical survival analysis and the causal estimands framework, there are no censoring tables. All explicit censoring for particular events commonly practiced in pharamaceutical clinical trials is replaced by an appropriate strategy to address them. Censoring is merely a catch-all addressing remaining loss to follow-up after appropriately accounting for intercurrent events. Under the FDA's current cancer endpoint guidance, however, traditional censoring tables are still generally expected.[11] We have accordingly attempted to translate between causal inference and conventional pharmaceutical terminology regarding censoring.

The estimands framework is not simply new language to describe conventional practices. It requires a rethinking of study conception, design, planning, execution, and analysis. We anticipate it will have impact on the data collection and interpretation of most if not all oncology studies with time-to-event analyses.

## 13. Data availability statement

No new data is presented in this manuscript. This manuscript is based solely on previously published results.